\documentclass{article}
\usepackage{amsmath}
\usepackage{amssymb}
\usepackage{xcolor}
\usepackage{graphicx}
\usepackage[round,authoryear]{natbib}
\setcitestyle{aysep={}}

\definecolor{wordbudget}{RGB}{0,102,153}
\newif\ifshowbudgetnotes
\showbudgetnotesfalse
\newcommand{\budgetnote}[1]{%
  \ifshowbudgetnotes
    \noindent\textcolor{wordbudget}{[#1]}\par
  \fi
}

\title{Optimal Transport in Economics}
\author{Amine Fahli and Alfred Galichon\\
Department of Economics, New York University\\
New York, NY, USA\\
\texttt{mxf5379@nyu.edu} (A.F.) \quad \texttt{ag133@nyu.edu} (A.G.)}
\date{October 2026}

\newcommand{\runningtitle}{Optimal Transport in Economics}

\begin{document}

\maketitle

\begin{center}
\small
ORCID: Alfred Galichon, 0000-0003-2495-8152\\
Corresponding author: Alfred Galichon, Department of Economics,
New York University, New York, NY, USA; \texttt{ag133@nyu.edu}\\
Running title: \runningtitle
\end{center}

\begin{abstract}
\budgetnote{approximately 150 words; outside main-text budget}
Optimal transport provides a common language for allocation, equilibrium,
computation, and inference.  Its primal problem assigns mass or agents, while
its dual variables admit economic interpretations as utilities, prices, and
scarcity rents.  This review explains why this combination has proved unusually
effective in economics.  We first present the core results, including
Kantorovich duality, integrality, cyclical monotonicity, the canonical distance
and quadratic costs, and entropic regularization.  We then trace the field's
development from planning and operations research to modern analysis,
statistics, and computation.  The economic literature is organized around two
roles for transport: as a model of matching, trade, hedonic equilibrium, and
aggregate assignment; and as a tool for coupling distributions, measuring
discrepancies, constructing multivariate ranks, solving inverse problems, and
certifying economic conclusions.  We conclude by examining extensions beyond
transferable utility, one-to-one matching, static allocation, and unconstrained
transport, together with open questions involving learning and identification
across multiple markets.
\end{abstract}



\section*{Significance statement}
\budgetnote{approximately 175 words; outside main-text budget}
Every economy must decide how to allocate scarce resources among people, firms,
and places.  Optimal transport turns this basic problem into a precise and
computable framework.  It connects the best feasible allocation to the prices
or rewards that can support it, and can accommodate both indivisible choices
and continuously distributed populations.  Economists use these ideas to study
matching in labor and marriage markets, consumer demand, international trade,
geographic inequality, and the effects of public policy.  The same mathematics
also helps researchers compare distributions, recover hidden preferences, and
compute counterfactual outcomes.  Recent analytical advances, scalable
algorithms, and increasingly data-rich allocation systems have made these
methods especially timely.  Understanding optimal transport therefore matters
both for explaining markets and for designing better ones.

We begin with the main optimal transport problems and results in a nutshell,
before placing their development in historical perspective.  We then explain
why optimal transport is so effective in economics and survey its principal
economic uses, first as a model and then as a tool.  The final section asks
where optimal transport stops, how its assumptions can be relaxed, and which
open questions appear most promising.

\section{Introduction}
\budgetnote{700 words}

Optimal transport is an unusually lively intellectual mixer, bringing
together mathematics, computation, operations research, probability, and
statistics, with economics near the center of the room.  Economics does more
than provide applications for optimal transport: it gives the theory's central
objects their most direct interpretation.  The primal problem allocates scarce resources;
the dual variables are utilities, prices, or scarcity rents; complementary
slackness identifies the pairs that trade; and strong duality is a welfare
theorem.  Convex analysis, linear programming, network algorithms, partial
differential equations, and statistical inference thus meet around questions
that are recognizably economic: who gets what, at what price, and what can be
learned from the resulting allocation.

A few connections convey the reach of the framework.  With only a change of
sign, essentially the same mathematical problem moves from complementarity
between workers and firms to substitution among products.  On the dual side,
utilities and prices bring matching, discrete choice, pricing, and market
equilibrium into a common framework.  Entropic regularization opens another
route: heterogeneous matching, structural gravity, and logit choice meet
matrix scaling and Sinkhorn's algorithm.  In continuous models, Brenier maps
supply multivariate ranks and quantiles, together with instruments for
identification and distributional comparison.  At the discrete end,
integrality accommodates indivisible agents and objects without leaving linear
programming.  Nor is this coherence confined to finite settings: under natural
conditions, the main conclusions survive the passage from discrete to
continuous marginals.  These connections play two roles throughout the
review.  Sometimes optimal transport is the model, with allocation and equilibrium as the
primal and dual problems.  Elsewhere it is a tool for identifying, estimating,
computing, or comparing objects generated by another economic model.

\begin{figure}[t]
\centering
\includegraphics[width=\textwidth]{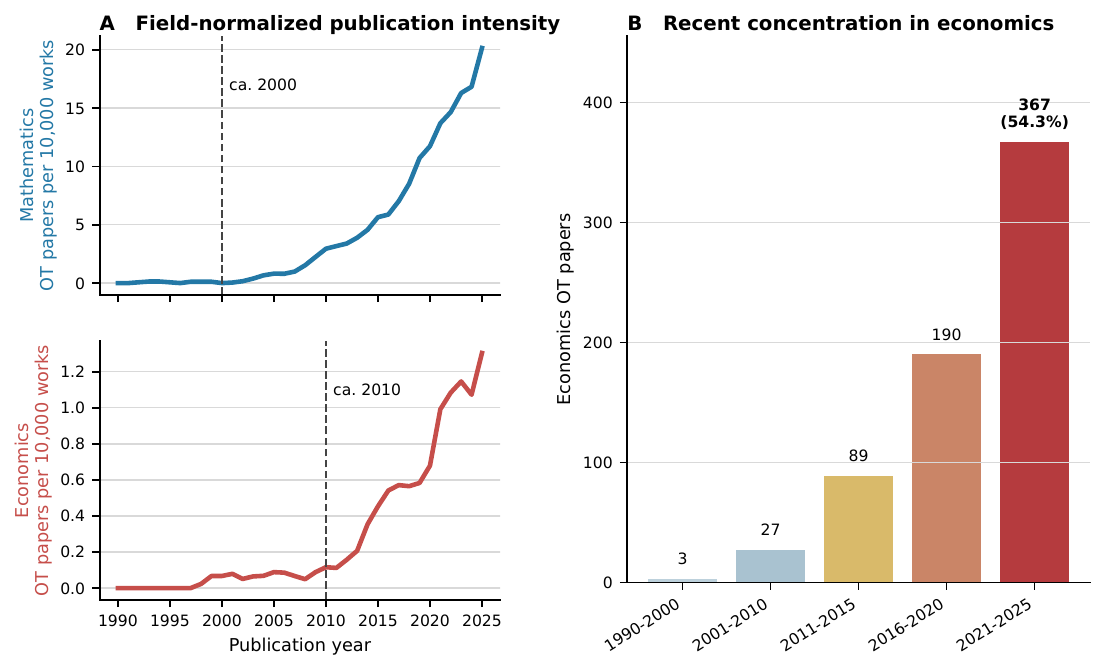}
\caption{The delayed takeoff and recent acceleration of optimal transport in
economics.  Panel A reports three-year trailing averages of papers containing
the exact phrase ``optimal transport'' per 10,000 works in mathematics and
economics.  The vertical lines mark the approximate breaks in the two
field-normalized trajectories.  Panel B reports economics papers by publication
period: 367 papers, or 54.3 percent of the 676 papers identified since 1990,
appeared during 2021--2025.  The corresponding mathematics sample contains
3,596 papers.  Fields are assigned by OpenAlex; the figure indicates publication
timing rather than providing an exhaustive census of optimal transport research.}
\label{fig:openalex}
\end{figure}

Why is this synthesis especially timely?  Three developments have converged.
First, the analytical theory has matured around results that economists can use
directly, from convex potentials and polar factorization to the geometry of
Wasserstein space.  Second, regularization and scalable algorithms have made
large transport problems routine rather than merely conceptual.  Third,
economic institutions increasingly do more than passively reveal equilibria:
they match participants, allocate capacity, set prices, and design markets,
while empirical work increasingly compares whole distributions rather than a
few moments.  The publication record mirrors this convergence.  In an OpenAlex
tabulation, the field-normalized mathematics trajectory turns upward around
2000, while economics follows roughly a decade later and accelerates sharply
after 2020.  More than half of all economics papers identified since 1990
appeared during 2021--2025 (Figure~\ref{fig:openalex}).

Our perspective is necessarily personal.  Optimal transport has occupied a
substantial part of one author's research for nearly two decades, first as a
language for matching and equilibrium and increasingly as a general method for
economic modeling and econometric inference.  This review is written for
economists familiar with standard graduate microeconomics and econometrics, but
assumes no previous exposure to optimal transport.  It complements three more
specialized treatments.  \citet{Galichon2016} develops the mathematical
foundations and their economic interpretation; \citet{Galichon2026} emphasizes
the particularly fruitful connection between optimal transport, random utility, and
entropy; and \citet{GalichonHenry2026Guide} provides a tool-oriented guide to
optimal transport in econometrics.  Here we organize the subject instead around
the economic ideas that account for its effectiveness, the areas in which it
operates as a model or as a tool, and the boundaries along which the framework
is now being extended.

\section{Optimal transport in a nutshell}
\budgetnote{2,100 words}

\subsection{The primal problem}
\budgetnote{250 words}

Begin with finite type sets $\mathcal X$ and $\mathcal Y$, carrying
nonnegative mass vectors $n=(n_x)$ and $m=(m_y)$ with equal totals.  A
transport plan is a nonnegative matrix $\mu=(\mu_{xy})$, where $\mu_{xy}$ is
the mass assigned from $x$ to $y$.  Given pairwise surplus $\Phi_{xy}$, the
primal problem is
\[
\max_{\mu\geq 0}\sum_{x,y}\mu_{xy}\Phi_{xy}
\quad\text{subject to}\quad
\sum_y\mu_{xy}=n_x,\qquad
\sum_x\mu_{xy}=m_y.
\]
This is the transportation linear program.  Its two sets of constraints ensure
that every unit on the first side is allocated and every capacity on the second
side is filled.

After normalizing total mass, the continuous counterparts are probability laws
$\mathcal L_X$ and $\mathcal L_Y$, and the matrix $\mu$ becomes a joint
distribution with these marginals.  Writing
$\Pi(\mathcal L_X,\mathcal L_Y)$ for the set of such couplings, the problem becomes
\[
\sup_{\mu\in\Pi(\mathcal L_X,\mathcal L_Y)}
\int\Phi(x,y)\,d\mu(x,y).
\]
Setting $c=-\Phi$ gives the equivalent cost-minimization formulation.  A
transport plan permits splitting, while a transport map is a pure allocation
$y=T(x)$, represented by
$\mu=(\operatorname{Id},T)_{\#}\mathcal L_X$, where
$T_{\#}\mathcal L_X=\mathcal L_Y$.  The measure $T_{\#}\mathcal L_X$, called
the pushforward of $\mathcal L_X$ by $T$, is simply the distribution of $T(X)$
when $X$ has distribution $\mathcal L_X$.  The existence of an optimal map is a
result, not part of the definition.

In matching, $\mu$ is the actual allocation of workers to firms or partners.
In discrete choice, it can be an analytical coupling between heterogeneous
tastes and chosen products, with the second marginal given by observed market
shares.  The same formulation therefore accommodates interactions between
complements and, after a change of sign, choices among substitutes.

\subsection{Kantorovich duality and the welfare theorem}
\label{sec:duality}
\budgetnote{400 words}

For the finite primal problem, associate potentials $u_x$ and $v_y$ with the
two sets of marginal constraints.  The dual problem is
\[
\min_{u,v}
\left\{
\sum_x n_xu_x+\sum_y m_yv_y:
u_x+v_y\geq \Phi_{xy}
\right\}.
\]
Every feasible pair $(u,v)$ places an upper bound on feasible total surplus.
Indeed, multiplying $u_x+v_y\geq\Phi_{xy}$ by any feasible $\mu_{xy}$ and
summing gives
\[
\sum_{x,y}\mu_{xy}\Phi_{xy}
\leq
\sum_xn_xu_x+\sum_ym_yv_y.
\]
Linear-programming duality says that the best such upper bound equals the
maximum surplus.  The difference between the dual minimum and the primal
maximum is called the duality gap, so this equality says that the gap is zero.
Moreover, both optima are attained: there exist a maximizing allocation and
minimizing potentials.

Complementary slackness sharpens this result:
\[
\mu_{xy}>0
\quad\Longrightarrow\quad
u_x+v_y=\Phi_{xy}.
\]
Thus the dual inequality is tight on every pair that is actually formed.
Conversely, any feasible allocation $\mu$ and feasible potentials $(u,v)$
satisfying equality $\mu$-almost everywhere are jointly optimal.  This is the
most elementary form of the welfare theorem contained in optimal transport.

In a transferable-utility matching market, $u_x$ and $v_y$ are equilibrium
payoffs.  The inequality $u_x+v_y\geq\Phi_{xy}$ says that no worker--firm pair
can generate enough surplus to make both members better off by deviating.
Equality on matched pairs says that their joint output is fully divided between
them.  The planner's surplus-maximizing allocation and the decentralized stable
allocation are therefore the primal and dual descriptions of the same outcome.
In other applications, the potentials are prices, scarcity rents, or value
functions.  They are also the shadow values of the marginal constraints, up to
the normalization $u\mapsto u+a$, $v\mapsto v-a$.

For general type spaces, the dual becomes
\[
\inf_{u,v}
\left\{
\int u\,d\mathcal L_X+\int v\,d\mathcal L_Y:
u(x)+v(y)\geq\Phi(x,y)
\right\}.
\]
If $\mathcal X$ and $\mathcal Y$ are compact metric spaces and $\Phi$ is
continuous, optimal plans and potentials exist, the primal and dual values
coincide, and complementary slackness holds $\mu$-almost everywhere.  More
general results replace compactness by tightness and suitable integrability
conditions, although dual attainment then requires care.  In finite dimensions
the result follows from linear-programming duality; in continuous settings it
follows through approximation, compactness, and the structure of transport
potentials.  The economic welfare logic survives the passage from finitely many
types to continua.

\subsection{Discrete transport and integrality}
\budgetnote{250 words}

For finite type spaces, the feasible optimal transport plans form a
transportation polytope, the set of nonnegative matrices satisfying the
prescribed row and column sums.  An extreme point is a feasible plan that
cannot be expressed as a nontrivial mixture of two other feasible plans.
Because the objective is linear, an optimum can always be selected among these
extreme points, which have considerable combinatorial structure.

The simplest case is the assignment problem, with the same number of
individual agents on each side and one unit assigned to each agent.  The
feasible plans are bistochastic matrices, meaning nonnegative matrices whose
row sums and column sums are all equal to one.  The
Birkhoff--von Neumann theorem states that every bistochastic matrix is a convex
combination of permutation matrices.  Consequently, although the linear
program permits fractional assignments, it always admits an optimal
deterministic matching.  Allowing mixtures or mass splitting does not improve
the optimum \citep{Birkhoff1946}.

The same conclusion extends beyond unit assignments.  The constraint matrix
of the transportation problem is the incidence matrix of a bipartite network
and is totally unimodular, meaning that every square submatrix has determinant
$0$, $1$, or $-1$.  Consequently, if the marginal masses $n$ and $m$ are
integers, every extreme point is integral, so an optimal plan can be found by
solving the continuous linear program rather than a generic integer program.
There is no integrality gap: relaxing the requirement of whole-agent
assignments does not improve the objective.  When $n_x$ represents several observationally identical agents,
integrality means that $\mu_{xy}$ counts whole agents; it does not require all
agents of type $x$ to receive the same assignment.

The network representation also explains computation.  Orient an arc from
every $x$ to every $y$, let $\mu_{xy}$ be its flow, and impose supplies $n_x$
and demands $m_y$ at the nodes.  Extreme plans are supported on acyclic
subgraphs, meaning collections of active links containing no cycle, and use at most
$\lvert\mathcal X\rvert+\lvert\mathcal Y\rvert-1$ positive arcs.  This
structure underlies assignment, transportation-simplex, and minimum-cost-flow
algorithms.  It is special, however: additional constraints can destroy total
unimodularity, at which point fractional solutions and genuine
integer-programming difficulties may reappear.

\subsection{Cyclical monotonicity}
\budgetnote{300 words}

Duality characterizes optimality through supporting potentials: dual payoffs or
prices satisfying $u(x)+v(y)\geq\Phi(x,y)$ for every possible pair, with
equality on the pairs used by the allocation.  Cyclical
monotonicity gives an equivalent condition directly on the pairs used by the
allocation.\footnote{In a quasilinear mechanism with gross utility
$\Phi(x,y)$, implementability by transfers is equivalent to
$\Phi$-cyclical monotonicity: transfers cancel when incentive constraints are
summed around a cycle \citep{Rochet1987}.  \citet{NoldekeSamuelson2018}
establish the general connection between stable matching and implementability,
which they call ``implementation duality.''}  A set
$\Gamma\subseteq\mathcal X\times\mathcal Y$ is
$\Phi$-cyclically monotone if, for every finite collection
$(x_i,y_i)\in\Gamma$,
\[
\sum_{i=1}^{k}\Phi(x_i,y_i)
\geq
\sum_{i=1}^{k}\Phi(x_i,y_{i+1}),
\qquad y_{k+1}=y_1.
\]
The left-hand side is the surplus generated by the observed pairs; the
right-hand side is the surplus after partners are reassigned around a cycle.
If the inequality failed, this reassignment would preserve both marginals
while increasing total surplus.  Every optimal coupling must therefore be
concentrated on a cyclically monotone set.

Under standard continuity and integrability conditions, the converse also
holds: a feasible coupling concentrated on a $\Phi$-cyclically monotone set is
optimal.  The link with duality is constructive.  Cyclical monotonicity permits
the construction of potentials $u$ and $v$ such that
\[
u(x)+v(y)\geq\Phi(x,y)
\]
for every pair, with equality on $\Gamma$.  Section~\ref{sec:duality} then certifies
optimality.  For general costs this construction uses $c$-convex potentials.
These are the cost-adjusted analogue of ordinary convex functions.  For
bilinear or quadratic surplus, Rockafellar's theorem says that a cyclically
monotone relation is contained in the subdifferential of a convex function, the
set of slopes supporting that function at a given point.  When the potential
is differentiable, this set contains a single gradient, which represents the
allocation and anticipates Brenier's theorem.

Economically, the cycle inequalities provide a test of revealed optimality.
Given an observed allocation and a proposed surplus function, they ask whether
any finite exchange of partners would raise aggregate output.  The supporting
potentials then rationalize the allocation as a stable equilibrium.  Because
the inequalities are pointwise in $x$ and $y$, they do not require specifying
either marginal distribution, making them useful for semiparametric
identification, estimation, and testing.  In one dimension, the two-pair
inequality already yields familiar sorting
predictions: supermodularity rules out crossed matches and produces positive
assortative matching, while submodularity reverses the pattern.  In several
dimensions there is generally no scalar ordering of types, but cyclical
monotonicity continues to characterize the geometry of optimal matches.

\subsection{Two canonical costs}
\budgetnote{500 words}

\subsubsection{Distance cost: duality and Beckmann flows}
\budgetnote{230 words}

When $x$ and $y$ represent locations and moving one unit costs its travel
distance, the natural cost is
\[
c(x,y)=\lVert x-y\rVert.
\]
This provides a benchmark for spatial allocation, shipping, and network-flow
problems.  For distributions with finite first moments, the optimal transport
value is the Wasserstein distance
\[
W_1(\mathcal L_X,\mathcal L_Y)
=
\inf_{\mu\in\Pi(\mathcal L_X,\mathcal L_Y)}
\int\lVert x-y\rVert\,d\mu(x,y).
\]
The triangle inequality allows the two Kantorovich potentials to be replaced
by a single function.  Kantorovich--Rubinstein duality gives
\[
W_1(\mathcal L_X,\mathcal L_Y)
=
\sup_{\operatorname{Lip}(p)\leq1}
\left\{
\int p(z)\,d\mathcal L_X(z)-\int p(z)\,d\mathcal L_Y(z)
\right\}.
\]
Here $\operatorname{Lip}(p)\leq1$ means that
$\lvert p(x)-p(y)\rvert\leq\lVert x-y\rVert$ for every $x$ and $y$.
The Lipschitz restriction has a natural price interpretation: spatial prices
cannot change faster than the cost of moving the good.

The same problem has a local formulation.  Instead of recording each origin
and destination through a coupling $\mu$, let $J$ be a vector-valued flow and
define the signed quantity measure $q=\mathcal L_Y-\mathcal L_X$.  At each
location $z$, $J(z)$ records the volume and direction of mass moving through
that location, much like a local shipment flow.  The Beckmann
formulation is
\[
W_1(\mathcal L_X,\mathcal L_Y)
=
\inf_{-\nabla\cdot J=q}
\int\lVert J(z)\rVert\,dz.
\]
For general measures, the objective is the total variation of the vector
measure $J$.  The divergence constraint imposes local conservation: flow is
created where $\mathcal L_X$ exceeds $\mathcal L_Y$ and absorbed where
$\mathcal L_Y$ exceeds $\mathcal L_X$.
Complementary slackness says that the Lipschitz constraint is saturated along
active transport directions.  Thus $W_1$ admits three equivalent descriptions:
a global coupling, a price potential, and a local flow.  On a discrete space,
the Beckmann formulation becomes a minimum-cost network-flow problem.

\subsubsection{Quadratic cost: Brenier's theorem and the Monge solution}
\budgetnote{270 words}

Quadratic cost has an immediate economic interpretation.  Because
\[
\frac12\lVert x-y\rVert^2
=
\frac12\lVert x\rVert^2+\frac12\lVert y\rVert^2-x\cdot y,
\]
and the first two terms are fixed by the marginals, minimizing quadratic cost
is equivalent to maximizing the bilinear surplus $x\cdot y$.  This makes the
problem directly relevant to assignment and matching.  We therefore take
\[
c(x,y)=\frac12\lVert x-y\rVert^2.
\]
If $\mathcal L_X$ and $\mathcal L_Y$ have finite second moments and
$\mathcal L_X$ is absolutely continuous, meaning that it has a density with
respect to Lebesgue measure, Brenier's theorem
\citep{Brenier1987,Brenier1991} states that the
optimal coupling is unique and induced by a map
\[
T(x)=\nabla\varphi(x),
\]
where $\varphi$ is convex and $T_{\#}\mathcal L_X=\mathcal L_Y$.  Thus, although
the optimization ranges over all couplings, including those that split mass,
its solution is pure: almost every $x$ is assigned to a single destination
$y=T(x)$.  This is what is traditionally called a Monge solution.  The proof
begins with cyclical monotonicity:
the support of the optimal plan is contained in the subdifferential of a convex
function.  Convex functions are differentiable almost everywhere, so the
subgradient is a unique gradient at $\mathcal L_X$-almost every $x$.

When $\mathcal L_X$ and $\mathcal L_Y$ have densities $f$ and $g$,
conservation of mass formally
yields the Monge--Amp\`ere equation
\[
f(x)
=
g(\nabla\varphi(x))
\det D^2\varphi(x).
\]
This converts the allocation problem into a nonlinear partial differential
equation.  Brenier's result also underlies polar factorization: a sufficiently
regular map can be decomposed into a measure-preserving rearrangement and a
monotone convex-gradient map.

Mapping a reference distribution to $\mathcal L_Y$
provides a multivariate quantile map; the inverse map provides multivariate
ranks.  Unlike $W_1$, which emphasizes local flows and Lipschitz prices,
quadratic optimal transport selects a canonical gradient map and exposes the convex
geometry of the allocation.

\subsection{Entropic optimal transport and Sinkhorn scaling}
\budgetnote{400 words}

For finite type spaces, entropy regularization replaces the linear primal
problem by
\[
\max_{\mu\in\Pi(n,m)}
\left\{
\sum_{x,y}\mu_{xy}\Phi_{xy}
-\varepsilon\sum_{x,y}\mu_{xy}\log\mu_{xy}
\right\},
\qquad \varepsilon>0.
\]
The marginal constraints remain exact.  Entropy changes only the objective,
penalizing concentrated plans and making the problem strictly concave.  With
positive marginals and finite surplus, the regularized problem has a unique,
strictly positive solution that varies smoothly with the surplus and the
marginals.
There is no scale-free default for $\varepsilon$: only
$\Phi/\varepsilon$ matters, and smaller values better approximate the
unregularized problem but worsen numerical conditioning.

Let $u_x$ and $v_y$ be the dual potentials.  The first-order conditions give,
after absorbing an additive constant into either potential, the Gibbs form
\[
\mu_{xy}
=
\exp\left(
\frac{\Phi_{xy}-u_x-v_y}{\varepsilon}
\right)
=
a_xK_{xy}b_y,
\qquad
K_{xy}=\exp\left(\frac{\Phi_{xy}}{\varepsilon}\right),
\]
where $a_x=\exp(-u_x/\varepsilon)$ and
$b_y=\exp(-v_y/\varepsilon)$.  In continuous formulations, $u$ and $v$ are
often called Schr\"odinger potentials \citep{Schrodinger1931,Leonard2014}.
Computing the optimal plan therefore
amounts to finding two scaling vectors such that $a_xK_{xy}b_y$ has row sums
$n$ and column sums $m$.

This leads to the alternating updates
\[
a_x\leftarrow
\frac{n_x}{\sum_yK_{xy}b_y},
\qquad
b_y\leftarrow
\frac{m_y}{\sum_xa_xK_{xy}}.
\]
The procedure is now called Sinkhorn or Sinkhorn--Knopp scaling
\citep{Sinkhorn1964,SinkhornKnopp1967}, but it is one
of the most frequently reinvented algorithms in science.  Other names include
iterative proportional fitting or IPFP, iterative proportional scaling,
matrix scaling or balancing, the RAS algorithm, raking, biproportional fitting
or scaling, the Deming--Stephan procedure, Kruithof's method, the Furness
procedure, and Bregman's procedure.  \citet{Idel2016} surveys this history.

The updates are elementary, but their convergence draws on deeper mathematics.
The scaling problem is a fixed-point problem for positive operators.
Perron--Frobenius theory provides the positive eigenvectors governing this
fixed point, while Birkhoff's theorem shows that the relevant positive map
contracts Hilbert's projective metric, a scale-invariant measure of
proportional differences between positive vectors
\citep{Birkhoff1957}.  For a
strictly positive finite kernel, the scaling vectors are unique up to a common
multiplicative normalization and the iterates converge.

As $\varepsilon\downarrow0$, the regularized value approaches the unregularized
transport value, and accumulation points of the regularized plans are optimal
transport plans.  When the original problem has several solutions, entropy
selects the one with maximal entropy.  Regularization thus connects a sparse
linear allocation problem to smooth matrix scaling.  Its speed, numerical
stability, and differentiability have made optimal transport practical at
large scales: specialized GPU implementations can compute gradients of
Sinkhorn losses for large empirical marginals \citep{PeyreCuturi2019}, while the
same balancing equations appear in structural gravity and other economic
models with two adding-up constraints.

\section{A historical perspective}
\budgetnote{1,200 words}

\subsection{A history of optimal transport}
\budgetnote{500 words}

A curious precursor stands outside the chronology. In work on differential
equations probably dating from 1836--1840, Jacobi devised a polynomial
assignment algorithm essentially equivalent to the later Hungarian method.
Published posthumously, it was identified by \citet{Ollivier2009} and discussed by
\citet{Kuhn2012}. The conventional history begins with Monge. In 1781, he asked how to move soil from
excavations (\emph{d\'eblais}) to fills (\emph{remblais}) at minimum
distance \citep{Monge1781}. His formulation assigned each particle to a destination,
requiring a map without mass splitting. The problem proved resistant.
Nineteenth-century French Academy of Sciences competitions produced
necessary conditions rather than existence theory; Appell's memoir, submitted
for the 1884 Bordin Prize competition and rewarded in 1885, treated continuous
and discontinuous systems but did not close the gap. \citet{Tolstoi1930}, working
for the Soviet transport administration, later
attacked cargo movements on railway networks. His cycle criterion and solution
of a 10-by-68 instance brought the problem close to modern network optimization.
These strands remained disconnected, awaiting the unifying language of linear
programming.

The breakthrough came in 1939 with Kantorovich's convexification of the
problem. Prompted by a problem in plywood production, he formulated linear programs
\citep{Kantorovich1939} and
interpreted his ``resolving multipliers'' as scarcity values. His 1942 paper on
the translocation of masses \citep{Kantorovich1942} replaced Monge's deterministic map by a nonnegative
transport plan with prescribed marginals. This relaxation is akin to allowing
mixed strategies: in the assignment case, a bistochastic plan is a convex
mixture of deterministic matchings. But the analogy has limits. A plan may
represent splitting or population heterogeneity, and integrality often
guarantees a deterministic optimum. The Kantorovich--Gavurin method of
potentials provided a primal--dual algorithm.

Parallel developments followed in the West. \citet{Hitchcock1941} formulated the
transportation problem; during 1942--1944 Koopmans helped allocate scarce Allied
merchant shipping across routes and cargoes, including Atlantic crossings
\citep{Koopmans1949}.
Dantzig's simplex method, developed in 1947 and later systematized by
\citet{Dantzig1963}, and its transportation specialization turned
the formulation into a practical computational tool. \citet{vonNeumann1953}
identified optimal assignment with a zero-sum hide-and-seek game, making the
link between linear-programming duality and minimax explicit. Kantorovich and
Koopmans shared the 1975 Nobel Prize for optimal resource allocation.

From the 1980s onward, the modern analytical theory gave optimal transport a tractable structure in
several dimensions. For squared Euclidean distance cost, Brenier
\citep{Brenier1987,Brenier1991} showed that, under standard regularity
conditions, the optimal allocation is generated by the gradient of a convex
potential. Economically, this means that a multidimensional assignment can be
pure and summarized by a single potential. Conservation of mass then links the
source and target densities through the Monge--Amp\`ere equation.
\citet{Ruschendorf1996} contributed to the theory of duality and cyclical
monotonicity for general costs, while \citet{GangboMcCann1996} extended the map-based theory beyond
squared distance.

A parallel computational development made optimal transport usable at scale.
Sinkhorn's matrix-scaling algorithm, reinterpreted as entropy-regularized
optimal transport, provides a fast approximation to large allocation problems
\citep{GalichonSalanie2009,Cuturi2013}. The syntheses of
\citet{Villani2003,Villani2009} helped unite these geometric, analytical, and
computational strands.

\subsection{Optimal transport in economics: from planning to equilibrium and back}
\budgetnote{500 words}

In its first incarnation, optimal transport was an allocation technology.
Kantorovich and Koopmans began from known quantities, capacities, and costs. The
planner directly chose feasible flows to maximize output or minimize resource
use. The primal variables were physical allocations; the dual multipliers were
scarcity prices, measuring marginal value of relaxing each constraint. Thus
linear programming supplied a plan and its decentralized interpretation. This
was a forward exercise: economic primitives entered the program and optimal
quantities and supporting prices emerged. It fit naturally with wartime
logistics, regulated industries, and economies organized through administrative
planning. Transport was therefore not chiefly a method for learning preferences
from observed market behavior. It was a practical language for deciding how
scarce resources should be used.

After the war, especially from the 1970s onward, economics moved away from
comprehensive planning. Competitive markets, presumed efficient at equilibrium,
became the benchmark; policy mainly corrected specific market failures. The
empirical task changed accordingly. Rather than take preferences and
technologies as known and compute an allocation, economists observed choices,
prices, matches, or market shares and worked backward to recover preferences,
technologies, and frictions. Revealed preference and structural econometrics
then used those estimated primitives to conduct counterfactuals. This inversion
of direction mattered for optimal transport. Its classical forward question, how should given
masses be allocated under a known objective, no longer described the
profession's central activity. Transport survived inside particular equilibrium
models, but largely disappeared as an explicit organizing language of economics.

Yet transport never entirely left economics. \citet{ShapleyShubik1971}
formulated transferable-utility matching as surplus maximization, with dual
variables representing agents' equilibrium payoffs. \citet{Becker1973}
selected a coupling of male and female characteristics that
maximized aggregate household output; supermodularity generated positive assortative
matching. In both cases, the planner's primal problem and the market's stability
conditions were two sides of the same linear-programming duality. Related
structures appeared in labor assignment, hedonic equilibrium, and trade models.
They were rarely presented as optimal transport: economists spoke of matching,
competitive assignment, or equilibrium. Transport persisted under economic
names and interpretations.

The present revival has a supply and demand side. On the supply side, Brenier's
theorem turned multivariate transport into a tractable map, while entropic
regularization and Sinkhorn iterations made large problems computationally
accessible. These advances allow economists to use optimal transport as a tool for constructing
ranks and quantiles, coupling distributions, identifying models, and computing
counterfactuals. On the demand side, important parts of modern economies now
resemble planners. Digital platforms do not merely observe decentralized
equilibrium: they set prices, rank alternatives, recommend partners, route
vehicles, and allocate scarce capacity. Ride-hailing platforms match riders and
drivers while adjusting prices; online dating platforms order and recommend
partners; school-choice and labor platforms implement explicit assignment and
priority rules. Their objectives and constraints may be imperfectly observed,
and participation, incentives, stability, and competition still matter.
Nevertheless, visible algorithms increasingly perform allocation, giving a new
technological form to the Lange--Hayek debate over whether coordination can be
computed centrally or must exploit dispersed knowledge
\citep{Lange1936,Hayek1945,Stiglitz1994}. AI-enabled platforms combine
centralized computation with decentralized information, making that old debate
newly practical. Optimal transport consequently returns not only as an
analyst's toolbox, but also as a model of actual institutions that directly
organize economic activity.

\subsection{Optimal transport across contemporary science}
\budgetnote{200 words}

Beyond economics, optimal transport is useful whenever distances between
outcomes matter for comparing distributions. In statistics and machine
learning, it compares samples, averages distributions through barycenters,
adapts predictions from one population to another, and helps train generative
models. In imaging, it measures spatial displacement rather than comparing
pixels one by one. In partial differential equations and stochastic control,
it describes the evolution of a distribution as a flow of mass. For economics,
these neighboring fields provide the multivariate geometry and scalable
algorithms used throughout this review.

\section{The relevance of optimal transport to economics}
\budgetnote{1,800 words}

\subsection{The ``unreasonable effectiveness'' of optimal transport}
\budgetnote{650 words}

\subsubsection{One sign change, two economic worlds}
\label{sec:sign-change}
\budgetnote{160 words}

The first source of the ``unreasonable effectiveness'' of optimal transport is
a simple change-of-sign trick.\footnote{The terminology and framing follow
\citet{Galichon2021}.  The title echoes Wigner's essay, ``The Unreasonable
Effectiveness of Mathematics in the Natural Sciences'' \citep{Wigner1960}.}  A matching
problem joins complementary agents:
workers and firms, spouses, or buyers and sellers create surplus only when
paired.  Orient every edge of the bipartite network from $x$ to $y$ and set
$c=-\Phi$.  On the combined node set $\mathcal Z=\mathcal X\sqcup\mathcal Y$,
define signed quantities and node prices by
$q=(-n,m)$ and $p=(u,-v)$.  Thus origins have $q(x)=-n_x$ and
$p(x)=u_x$, while destinations have $q(y)=m_y$ and $p(y)=-v_y$.
The marginal constraints become network clearing, and the matching dual
inequality becomes the Kantorovich price inequality
$p(y)-p(x)\leq c(x,y)$.  The same problem can then be read as traders choosing
optimal routes between origins and destinations.  The function $p$ gives the
price of the good at every node, and goods at different locations are
imperfect substitutes \citep{Galichon2021}.  Complementarity across the two sides
has become gross substitutability within the transformed demand system, yielding
monotone comparative statics, lattice structure, and convergent coordinate
updates, including IPFP and Sinkhorn, even with high-dimensional types and
indivisible allocations.

\subsubsection{The welfare theorem in its purest form}
\budgetnote{170 words}

The first welfare theorem states that a competitive equilibrium is Pareto
efficient.  In the transferable-utility economies represented by optimal transport, the
conclusion is sharper: the equilibrium allocation maximizes utilitarian
welfare, with every agent receiving weight one.  Transfers determine how the
surplus from a match is divided, but cancel when individual utilities are
added, leaving aggregate welfare equal to total match surplus.  Equivalently,
in Negishi's representation of competitive equilibrium, welfare weights are
inversely proportional to the marginal utility of income.  Quasilinearity in
the transferable numeraire makes this marginal utility identical across
agents, so all welfare weights can be normalized to one.

Kantorovich duality makes this result immediate.  Dual feasibility implies
that the aggregate value of any system of supporting payoffs bounds the
surplus attainable under every feasible allocation.  Complementary slackness
makes this bound exact for the equilibrium allocation.  The competitive
allocation is therefore not merely immune to Pareto improvements: it solves
the equal-weight utilitarian planner's problem.  Conversely, every
surplus-maximizing allocation admits supporting dual payoffs.  This stronger
conclusion applies to competitive equilibria possessing the
transferable-utility transport structure, not to competitive equilibria in
general.

\subsubsection{Linear program, nonlinear economics}
\budgetnote{160 words}

Linearity applies to the transport plan, not to the economic predictions.
In Becker's one-dimensional assignment model, the sign of the cross-partial
determines positive or negative assortative matching: complementary types
sort together, whereas substitutes sort oppositely \citep{Becker1973}.  Yet the
matching map changes nonlinearly with distributions and technology.  The
supporting wage is an envelope of match surplus; under standard
complementarity and curvature conditions, its slope rises with skill,
generating a convex wage schedule and differential rents
\citep{Sattinger1979}.  \citet{Sattinger1993} surveys the broader assignment
approach.  This amplification connects assignment models to superstar effects
and executive compensation: small talent differences can support large
earnings differences when higher talent is assigned to larger markets or
firms \citep{Rosen1981,Tervio2008,GabaixLandier2008}.  Thus the same linear program
determines not only allocation but the distribution of earnings.  Changes
in skill supplies, task demand, firm size, or complementarities reshape both
sorting and wage dispersion, sometimes sharply
\citep{CostinotVogel2010,Lindenlaub2017}.  In continuous quadratic models, this nonlinearity is
encoded in a convex potential and the Monge--Amp\`ere equation.  Optimal transport thus
combines exact, scalable computation with rich nonlinear comparative
statics.

\subsubsection{Discrete and continuous economies in one language}
\budgetnote{160 words}

The same formulation covers economies with finitely many indivisible agents
and economies described by continuous distributions of types.  In the
finite case, a coupling is a nonnegative assignment matrix whose row and
column sums equal the two populations; in the continuous case, it is a
joint measure with prescribed marginals.  The primal and dual programs,
their economic interpretations, and strong duality remain essentially
unchanged, with integrals replacing sums.
As the total-unimodularity result above shows, integer marginals admit an
integral optimum, so indivisibilities need not be smoothed away.  In nonatomic
economies, a transport plan allows types to split across destinations, whereas
a transport map represents a pure assignment.  More broadly, optimal transport
provides a canonical example in which results from finite-dimensional
linear programming with discrete marginals extend, under natural
conditions, to the infinite-dimensional setting with continuous marginals.
Additional structure on the cost can ensure that the optimal plan is
induced by a map.

\subsection{Optimal transport and its toolboxes}
\budgetnote{500 words}

Optimal transport draws its effectiveness from three complementary mathematical toolboxes.
Optimization and computation tell us how to solve the allocation problem.
Convex analysis, geometry, and PDEs tell us how to characterize continuous
solutions.  Order, lattice, and exchange methods tell us what qualitative
predictions can be obtained without solving the model fully.  The progression
is therefore from computation, to characterization, to comparative statics.
These are not separate collections of techniques, but different views of the
same allocation problem.

\subsubsection{Optimization and computation}
\budgetnote{155 words}

Optimal transport is computationally attractive because several algorithmic viewpoints
reinforce one another.  In finite settings it is a linear program, but also a
minimum-cost flow problem on a bipartite network.  This network structure
yields specialized assignment and minimum-cost-flow algorithms tailored to
the marginal constraints.

Entropic regularization supplies a scalable counterpart to exact network
optimization.  The Sinkhorn/IPFP updates introduced above are coordinate
adjustments of the dual prices.  Their convergence is not merely numerical
luck: positive-operator and Perron--Frobenius theory, together with Birkhoff
contraction, explain why the iterates stabilize.  As regularization vanishes,
the smooth solution reconnects with the original transport problem.  Thus optimal transport
combines exact optimization, combinatorial structure, scalable approximation,
and a convergence theory reaching far beyond optimization.

\subsubsection{Convex analysis, geometry, and partial differential equations}
\budgetnote{155 words}

The squared-distance result introduced above illustrates how optimal transport connects
several analytical toolboxes.  Duality produces a convex potential, its
gradient describes the allocation, and conservation of mass yields a nonlinear
PDE.  For general costs, $c$-convex potentials and their contact sets play the
analogous role of certifying optimality.

When distributions have densities, the resulting PDE is the Monge--Amp\`ere
equation.  Its nonlinear solution
describes equilibrium assignment; its elliptic linearization describes how
assignment and dual prices respond to small changes in primitives, with the
Laplacian as the basic benchmark.  For quadratic cost, the Benamou--Brenier
formula represents squared Wasserstein distance as the minimum kinetic energy
along density paths satisfying the continuity equation, thereby characterizing
geodesics in Wasserstein space \citep{BenamouBrenier2000}.  Many evolution
equations can then be viewed as gradient flows, or steepest-descent dynamics for
objectives defined over distributions \citep{JKO1998,Otto2001}.
\citet{SteinerbergerTsyvinski2019} apply this perspective to continuous reform
of the income-tax schedule.

\subsubsection{Order, lattices, and exchange structures}
\budgetnote{115 words}

Order enters optimal transport through its dual problem, not through supermodularity of the
surplus.  After the change of sign in Section~\ref{sec:sign-change}, feasible node prices
satisfy $p(y)-p(x)\leq c(x,y)$.  These difference constraints define a lattice
under the componentwise order on $p$, while the linear dual objective
$\langle q,p\rangle$ is modular on this lattice.  Consequently, optimal node
prices form a lattice; translated back to $(u,v)$, the order on the destination
side is reversed.  Topkis-type arguments then yield comparative statics
\citep{Topkis1978}.  Under entropy
regularization, the constraints enter the objective, which becomes a smooth
submodular function.  Its gradient satisfies gross substitutes, yielding
inverse isotonicity and convergence of coordinate updates
\citep{GalichonSamuelsonVernet2025}.  $L$- and
$M$-convexity and exchange structures provide discrete-convex versions of the
same principle \citep{Murota2003}.  Separately, Rheinboldt's $M$-functions
combine off-diagonal antitonicity with inverse isotonicity and provide
convergence results for nonlinear coordinate iterations \citep{Rheinboldt1970}.

\section{Optimal transport in economics: a review}
\budgetnote{3,150 words}

The review that follows is organized around a distinction between optimal
transport as a model and optimal transport as a tool.  In the first case, the
optimal transport plan is an actual economic allocation, its marginal constraints
express feasibility or market clearing, and its dual potentials represent
prices, utilities, or scarcity rents.  In the second, optimal transport is introduced
by the analyst to couple distributions, construct statistical objects, identify
or estimate parameters, or certify optimality; the resulting potentials need
not correspond to market prices.  Some of the papers reviewed below formulate
this structure explicitly as optimal transport.  In other cases, the authors
do not use that terminology and the optimal transport interpretation is ours;
we flag these cases rather than attributing the formulation to the original
paper.  The boundary is deliberately porous.  In inverse optimal transport,
for example, optimal transport describes the allocation mechanism in the
forward direction and becomes an econometric tool when the mechanism is
inverted.  Book-length treatments of these connections are provided by
\citet{Galichon2016,Galichon2026}.

\subsection{Optimal transport as a model}
\budgetnote{1,750 words}

When optimal transport is the model, the coupling is the economic allocation
to be explained, while the dual potentials provide the prices or payoffs that
decentralize it.  This interpretation is clearest in matching and hedonic
markets: once the populations on two sides are fixed, the primal problem
determines who interacts with whom, while the dual variables determine wages,
profits, or utilities.  Optimal transport thereby connects sorting, welfare,
and equilibrium prices.  We review this role in matching, trade, and hedonic
models, including both the forward determination of allocations and the
recovery of primitives from observed outcomes.

\subsubsection{Matching: the forward problem}
\budgetnote{550 words}

\paragraph{Two margins: existence, equilibrium, and scarcity rents}

Matching is the canonical case in which optimal transport is the economic model itself rather
than a method imposed by the analyst.  With transferable utility, the
distributions of agents on both sides are fixed and a match produces a given
surplus.  \citet{Becker1973} studies the resulting sorting problem, while
\citet{ShapleyShubik1971} show that the core of the finite assignment game is exactly
the solution set of the dual assignment program.  Neither paper uses modern
optimal transport terminology, but their formulations have its primal--dual
structure.  In optimal transport terms, the matching is the primal coupling,
equilibrium payoffs are the dual potentials, and complementary slackness is the
no-blocking condition.  Stability and efficiency coincide
because transfers allow each pair's surplus to be divided freely.

\citet{GretskyOstroyZame1992,GretskyOstroyZame1999} extend this equivalence to
nonatomic assignment economies and give dual uniqueness a deeper economic meaning.  The
transport value, viewed as a function of the two populations, is a social-gains
function; its derivatives are the agents' equilibrium payoffs.
Differentiability of this value, uniqueness of the dual potentials, payment at
marginal product, and perfect competition become different statements of the
same property.  Nonuniqueness of potentials is therefore economically
meaningful: it may reflect rents and imperfect competition rather than a
technical nuisance.

\paragraph{Beyond classical sorting}

The classical one-dimensional model asks whether matching is positively or
negatively assortative.  Modern optimal transport matters because this dichotomy rarely
exhausts economically interesting couplings.  With multidimensional types on
both sides of the market, assortative matching is more complex.
\citet{Lindenlaub2017} provides a natural definition and derives a tractable
assignment that depends on the matrix of complementarities.  Applying this
framework to technological change, she uses worker--job data to estimate rising
cognitive and declining manual complementarities, significantly affecting
sorting and wage inequality.

When the dimensions of heterogeneity differ across the two sides, ordinary
monotonicity is not even defined.  \citet{ChiapporiMcCannPass2016,ChiapporiMcCannPass2017} show
that the multidimensional side is partitioned into level sets matched to the
lower-dimensional side.  Under their nestedness condition, these level sets
are ordered and the assignment can be constructed sequentially.  Nestedness
depends jointly on the surplus and the marginals, unlike supermodularity, which
is a property of surplus alone.

Team formation introduces more than two marginals.  \citet{CarlierEkeland2010}
establish the multi-marginal equilibrium formulation, while
\citet{BoermaTsyvinskiZimin2025} use it to characterize teams of several workers matched with
heterogeneous firms.  The solution combines mixed regions with countermonotone
regions, and the dual potentials become wages and firm values.  Mismatch costs
that are concave in the mismatch distance produce a different departure.  In
\citet{BoermaTsyvinskiWangZhang2023}, fixed
investments make additional mismatch progressively less costly, yielding
composite rather than positive or negative sorting.  Here optimal transport is not merely
convenient terminology: optimal transport with distance-concave mismatch costs supplies
a nested, hierarchical structure that can be exploited to characterize and compute an
allocation that standard assortative-matching arguments cannot describe.  Related
tractable problems with surplus that is neither supermodular nor submodular arise
in redistricting with endogenous candidates and nondiscriminatory personalized
pricing; their solutions combine assortative and counterassortative regions
\citep{Moscariello2026,StrackYang2026}.

\paragraph{When matching is not automatically optimal transport}

Transferable utility supplies both a scalar surplus and a welfare theorem.
General nontransferable-utility matching need not have either, so stability
should not automatically be represented as fixed-surplus optimal transport.
The equilibrium-transport framework discussed in
Section~\ref{sec:equilibrium-transport} handles this more general case by
replacing fixed-surplus maximization with pairwise feasibility conditions.
Here we focus instead on two special routes that recover optimal transport
structure.  First, under logit heterogeneity, random-utility aggregation
can generate a potential even when individual utilities are nontransferable.
The multiplicatively separable matching equations derived by \citet{Dagsvik2000},
and given a large-market foundation by \citet{Menzel2015}, can be read as the
first-order conditions of entropically regularized optimal transport.  Smoothing restores at
the aggregate level the symmetry absent from the individual problem.

Second, aligned preferences can embed stability in a family of transport
objectives.  \citet{GalichonGhelfiHenry2023} show that stability may force
extreme inequality; \citet{EcheniqueRootSandomirskiy2025} explain this
result by making stable matching a limit of transport problems with an
Atkinson-style cost that is concave in match utility, and trace the tension
between stability, utilitarian welfare, and egalitarianism.  This is a special
transport representation, not a general NTU welfare theorem.  Other departures
break the framework altogether.  When
match surplus depends on the allocation itself, through congestion, search, or
market power, the objective is no longer linear in a coupling.
\citet{LindenlaubOhPeters2025} provide a useful boundary case: firms sort across
locations in a pure manner, even though they subsequently hire workers in a
frictional labor market.

\subsubsection{Matching: identification and estimation}
\budgetnote{600 words}

\paragraph{From matching to surplus}

The forward matching problem takes the surplus matrix $\Phi$ as given and
predicts an equilibrium coupling $\mu$.  Empirical work reverses this map:
from the observed distribution of matches, what can be learned about the
surplus that generated it?  Without additional structure, the answer is very
little.  An unregularized optimal assignment is typically unchanged under
many perturbations of $\Phi$, so observing its support delivers inequalities
rather than point identification.  Unobserved heterogeneity changes this
conclusion.  It smooths individual choices and strictly convexifies the
aggregate matching problem, making the map from systematic surplus to matching
patterns invertible.

\citet{Graham2011} surveys econometric approaches to assignment with
complementarities and spillovers.  A complementary, distribution-free branch
uses rank-order restrictions to identify and estimate matching games
\citep{Fox2010,Fox2018}.

\citet{ChooSiow2006} provide the canonical example.  In their marriage model,
individuals within each observed type have idiosyncratic logit preferences over
partner types and over remaining single.  The resulting equilibrium is,
retrospectively, an unbalanced entropically regularized transport problem.
Consequently, the systematic surplus is recovered directly from the matching
table:
\[
 \Phi_{xy}
 =2\log\mu_{xy}-\log\mu_{x0}-\log\mu_{0y},
\]
where $\mu_{xy}$ denotes marriages between types $x$ and $y$, while
$\mu_{x0}$ and $\mu_{0y}$ denote singles.  Thus match frequencies substitute
for unobserved transfers.  \citet{GalichonSalanie2022} extend this argument
beyond logit heterogeneity.  Separability produces a generalized entropy
determined by the taste-shock distributions; convex duality then supplies both
the equilibrium computation and the inverse map from $\mu$ to $\Phi$.  In this
setting, regularization is not added for numerical convenience.  It is the
aggregate representation of economically meaningful heterogeneity.

\paragraph{What the data reveal}

This perspective also clarifies the informational content of different data.
Observing the primal object, the complete matching table including unmatched
agents, can identify systematic match surplus under the maintained
heterogeneity structure.  Observing only dual objects, such as wages, is
generally less informative.  \citet{EeckhoutKircher2011} show that wage data
may identify the strength of sorting while failing to identify its sign.
\citet{EkelandGalichon2013} develop the complementary, assumption-light
approach: cyclical monotonicity characterizes whether observed assignments and
prices can be rationalized, but generally identifies a set of admissible
preferences rather than a unique surplus.  In additive random-utility models,
the map from mean utilities to choice probabilities is cyclically monotone
without specifying the distribution of utility shocks, an insight exploited
by \citet{AllenRoyRehbeck2019,ShiShumSong2018,MeloPogorelskiyShum2019}.  These
results locate the central trade-off in inverse transport.  Distributional assumptions and strict
convexification yield a sharp inverse, while weaker assumptions yield robust
but set-valued conclusions.

Parametric and regularized versions turn this identification argument into
practical estimators.  With a bilinear surplus
$\Phi(x,y)=x^{\mathsf T}Ay$, the affinity matrix $A$ measures which attributes
are complements or substitutes across the two sides.  \citet{DupuyGalichon2014}
use its singular vectors to extract the combinations of traits most
relevant for sorting, and its rank to measure the effective dimensionality of
the market.  Low-rank penalties extend this idea to settings with many
characteristics \citep{DupuyGalichonSun2019}, while
\citet{LiYeZhouZha2019} develop nonlinear inverse-transport methods for noisy or incomplete
matching tables.

\paragraph{Counterfactuals and policy}

Once the surplus and heterogeneity have been recovered, counterfactuals have a
particularly transparent structure.  Changes in population composition move
the marginals; technological, institutional, or preference changes move the
surplus or the regularizer.  Resolving the transport problem separates these
channels and predicts their effects on matching, transfers, and welfare.  This
logic has been applied to taxation in matching markets
\citep{DupuyGalichonJaffeKominers2020}, educational expansion and wage premia
\citep{Corblet2026}, and long-term-care matching \citep{Gorlach2025}.
\citet{HazardKitagawa2025} instead treat
the coupling itself as a policy learned from data and provide welfare-regret
bounds that account for estimation error.  \citet{SunadaIzumi2025} formulate
capacity-constrained treatment assignment as an optimal transport problem.
With continuous covariates or treatments, the Wasserstein metric makes the
planner's space of feasible joint distributions compact and provides a unified
framework for local asymptotic analysis; with finite covariates and binary
treatment, the optimal transport formulation is largely a matter of
representation.  The general contribution is not
any one application: inverse optimal transport turns equilibrium sorting from an outcome to
be described into a source of information about complementarities and a
platform for policy counterfactuals.

\subsubsection{Trade and hedonic models}
\budgetnote{350 words}

\paragraph{Hedonic models}

The matching surplus need not be primitive.  In a hedonic market, consumer
$x$ and producer $y$ interact through a quality $z$.  If $U(x,z)$ is utility
and $C(y,z)$ is production cost, optimizing over quality yields
\[
\Phi(x,y)=\sup_z\{U(x,z)-C(y,z)\}.
\]
The product is chosen within the pair, while optimal transport determines which
consumer and producer types are paired.  This construction therefore reduces a
richer market to an assignment problem.  \citet{Ekeland2010} proves existence,
uniqueness, and efficiency of hedonic equilibrium.  \citet{ChiapporiMcCannNesheim2010}
establish the equivalence between hedonic equilibrium, stable matching, and
optimal transport: the equilibrium coupling is the primal allocation and the
nonlinear product-price schedule is reconstructed from the dual potentials.

This equivalence also organizes identification.  \citet{EkelandHeckmanNesheim2004}
show that preferences and technology can be identified from a
single market by exploiting the restrictions encoded in the nonlinear
equilibrium price schedule.  \citet{ChernozhukovGalichonHenryPass2021}
extend this logic to multiattribute products and multidimensional unobserved
heterogeneity, using generalized convexity and optimal transport to identify nonseparable
simultaneous-equation systems.

\paragraph{Trade, gravity, and space}

Trade connects to optimal transport at two levels of aggregation.  Spatial
models describe how goods are routed through a network, with edge flows
satisfying conservation at every location; this leads to Beckmann's flow
formulation.  Congestion and infrastructure choices may make route costs depend
on the allocation itself, with optimal transport providing the benchmark from
which these richer models depart.  Gravity models instead describe total
shipments between origins and destinations, whose margins are output and
expenditure.  Under common gravity structures, bilateral flows have the
multiplicative form of an entropically regularized optimal transport solution.
The origin and destination scaling factors are dual potentials, and Sinkhorn
scaling adjusts them until both margins clear.

\citet{KoopmansBeckmann1957} show that assigning activities to locations is a
linear transport problem supported by land rents.  Interactions among plants,
however, make the objective quadratic and destroy this decentralization.
\citet{FajgelbaumSchaal2020} use Beckmann-type flows to embed the routing of
goods and the choice of infrastructure in spatial equilibrium.

Structural gravity, surveyed by \citet{HeadMayer2014} and including
\citet{AndersonVanWincoop2003}, writes bilateral
trade as $X_{ij}=a_iK_{ij}b_j$, with the exporter and importer factors enforcing
the two margins.  In optimal transport terms, this system is an entropically
regularized problem: multilateral-resistance terms are dual
potentials and their computation is Sinkhorn scaling; the corresponding
existence, uniqueness, and computation problem is developed by
\citet{AllenArkolakisTakahashi2020}.  \citet{EatonKortum2002}
derive a related gravity system from Ricardian discrete choice rather than
two-marginal assignment.  Finally, the fixed-effect first-order conditions of
PPML gravity estimation \citep{SantosSilvaTenreyro2006} reproduce the observed
margins and can be read as an inverse entropic optimal transport problem.

\subsubsection{Macroeconomics: assignment inside aggregate equilibrium}
\budgetnote{250 words}

Macroeconomic models rarely end with the solution of a transport problem.
Conditional on distributions of workers, jobs, tasks, firms, or locations, optimal transport
determines assignment and the associated wages or scarcity rents.  Those prices
then affect occupational choice, investment, migration, firm size, and
accumulation, making the marginals or the surplus endogenous.  The complete
model therefore embeds an optimal transport block inside a general-equilibrium fixed point or
a dynamic system.  This decomposition separates comparative-advantage
assignment from the forces governing aggregate quantities and their evolution.

\citet{Sattinger1979} interprets wages as differential rents generated by the joint
scarcity of workers and jobs.  \citet{CostinotVogel2010} embed
comparative-advantage assignment in general equilibrium and connect changes in
technology and factor supplies to inequality.  With multidimensional skills and
tasks, \citet{Lindenlaub2017} shows how changing complementarities reshape both
sorting and the wage distribution.  \citet{BojilovGalichon2016} obtain
closed-form equilibrium and comparative statics in a quadratic-Gaussian
matching model with taste heterogeneity.  More generally,
\citet{BoermaOttoliniTsyvinski2025} decompose the response to technological
change into a gradient component, which determines changes in marginal earnings
through a Poisson equation, and a divergence-free component, which determines
labor reallocation.  Team production and firm organization
introduce additional margins or endogenous group composition, as in
\citet{BoermaTsyvinskiZimin2025} and the composite-sorting model of
\citet{BoermaTsyvinskiWangZhang2023}.

The second marginal is essential.  When occupational supplies adjust freely,
workers may simply select activities according to absolute advantage, as in the
Roy interpretation of \citet{HsiehHurstJonesKlenow2019}.  Similarly,
\citet{Gola2021} combines Roy-style selection across sectors with Becker-style
matching within sectors: the within-sector assignments can be read as
optimal transport blocks, while sectoral labor supplies are determined jointly
in equilibrium.  More generally, optimal transport enters
when capacities, job distributions, or other aggregate feasibility conditions
link individual choices.  The same qualification applies to public allocation:
fixed capacities create a transport problem only when the mechanism also
supplies an objective or equilibrium condition selecting among feasible
assignments.

\subsection{Optimal transport as a tool}
\budgetnote{1,400 words}

When optimal transport is a tool, the analyst introduces it to answer a
particular question rather than to describe the economic allocation itself.  In
econometrics, it can characterize all couplings compatible with observed
marginals and thereby produce sharp bounds; select a canonical map to construct
multivariate ranks, quantiles, or identifying restrictions; or supply a
distance, estimation criterion, or robustness neighborhood for comparing model
and data.  When the analyst selects a particular map or metric, the implied
geometry is not necessarily dictated by the economic model and therefore
requires justification.  The applications below are organized by what optimal
transport is asked to do: couple distributions, measure discrepancies,
construct ranks and quantiles, invert models, or certify and characterize
solutions.

\subsubsection{Couple}
\budgetnote{280 words}

Many econometric problems identify marginal distributions but leave their
dependence unknown.  The admissible joint distributions are then precisely the
couplings $\Pi(\mathcal L_X,\mathcal L_Y)$.  Maximizing and minimizing a parameter over this set
produces sharp bounds.  Fr\'echet--Hoeffding bounds are the scalar prototype
\citep{Hoeffding1940,Frechet1951};
optimal transport extends the same logic to general objectives, conditional distributions,
support restrictions, and multiple marginals.  Here transport does not model
an allocation.  It exhausts the dependence structures that the available
information has not ruled out.

\citet{GalichonHenry2011} apply this principle to incomplete structural models
with multiple equilibria.  A parameter is observationally admissible exactly
when the observed outcome distribution can be coupled with latent variables on
the graph of the model's correspondence.  A zero-one transport cost turns this
support condition into a scalar specification criterion, while duality yields
a Strassen-type characterization \citep{Strassen1965}.
\citet{DaljordHuPouliotXiao2019} use the same
logic to infer otherwise unobserved trade in Beijing's black market for license
plates: the minimum mass that must be transported between the pre- and
post-rationing car-price distributions gives a sharp lower bound on the volume
of illegal transactions.  \citet{DHaultfoeuilleGaillacMaurel2025} use
Strassen's theorem and the radial function to obtain sharp identified sets when
variables observed in separate datasets cannot be linked.  In causal inference,
\citet{FanGuerreZhu2017,Russell2021} bound functionals of the
unobserved joint distribution of potential outcomes.  \citet{JiLeiSpector2023}
use Kantorovich duality to unify and extend a broad class of partially identified
causal estimands.  Any feasible estimated dual potentials deliver valid bounds,
so first-stage errors affect sharpness without invalidating coverage; the dual
formulation also yields double-machine-learning-style second-order accuracy,
including with continuous potential outcomes.

Other applications move from bounding the missing coupling to selecting or
estimating one.  \citet{ArellanoBonhomme2012,ArellanoBonhomme2017} recover latent
heterogeneity or dependence in deconvolution and selection models.
More directly, \citet{ArellanoBonhomme2023} use optimal transport matching to
estimate linear models with independent latent variables nonparametrically.
\citet{GunsiliusXu2022} use multimarginal unbalanced optimal transport to match
covariates simultaneously across multiple treatment arms.
\citet{RigolletWeed2018} show that entropic optimal transport with quadratic cost
corresponds to maximum-likelihood deconvolution under Gaussian noise when the
model class is closed under domination.  Their relaxed formulation covers the
more general case in which this closure condition does not hold.
\citet{TorousGunsiliusRigollet2024} impose the Brenier coupling as the multivariate analogue of
the monotone changes-in-changes coupling of \citet{AtheyImbens2006}.  The distinction is essential: optimal transport may either
describe every dependence pattern compatible with the data, or impose one
canonical coupling as an identifying restriction.

\subsubsection{Measure}
\budgetnote{300 words}

Optimal transport can also be chosen by the analyst as a geometry for comparing
distributions.  This differs from the coupling problems above: the unknown
dependence structure is no longer the object of interest.  Instead,
Wasserstein distance measures how much, and how far, probability mass must be
moved to transform one distribution into another.  The ground cost is therefore
substantive.  It determines which discrepancies the procedure regards as large
and must be justified in terms of the economic variables.

One use is to construct averages of distributions.  \citet{Gunsilius2023} defines a
distributional synthetic control as a Wasserstein barycenter of the control
units, with weights chosen to reproduce the treated unit before treatment.  On
the line, this barycenter is simply a weighted average of quantile functions,
so the method extends synthetic controls from means to entire outcome
distributions.  \citet{GunsiliusHsiehLee2024} extend this approach to
multivariate outcome distributions using tangential Wasserstein projections,
which also provide a general notion of projection in Wasserstein space.  A
second use is to measure distance from a model.
\citet{SchennachStarck2026} choose the smallest Wasserstein perturbation of the empirical
distribution that satisfies the model's moment conditions.  Optimally
transported GMM moves observations rather than reweighting them; its value
measures the specification or measurement error needed to reconcile model and
data, while its transport map shows where that adjustment occurs.

Distributional robustness reverses this minimization.  Rather than finding the
smallest perturbation that makes a model valid, it finds the worst outcome
among distributions within a fixed Wasserstein budget.  Transport duality makes
these problems tractable \citep{BlanchetKangMurthy2019}, while
\citet{AdjahoChristensen2022} use the approach to choose treatments that remain effective
when the target population differs from the experimental population.  The
ability to move mass outside the observed support is crucial for external
validity.

Finally, an entropic transport model can provide a structured benchmark.
Recent work on the international trade network compares observed flows with a
gravity-based optimal transport null and detects communities in the residual network
\citep{MastrandreaPagnottoniPecoraSpelta2025}.  This
is an adjacent network-science application rather than a structural welfare
model, but it illustrates the general principle: optimal transport may supply a mean, a
distance, an ambiguity set, or a disciplined null model.

\subsubsection{Rank and sort}
\budgetnote{260 words}

On the line, the distribution function ranks observations, its inverse produces
quantiles, and monotone rearrangement couples two distributions.  These are
three aspects of the same ordering.  In several dimensions there is no
canonical order, but quadratic optimal transport provides a replacement.
\citet{ChernozhukovGalichonHallinHenry2017} fix a reference distribution $\nu$ and
define the vector rank of an outcome $Y$ with law $\mathcal L_Y$ as the Brenier
map $R$ transporting $\mathcal L_Y$ to $\nu$.  Its inverse $Q$ is the vector
quantile map.  The norm of $R(y)$ measures
outlyingness, its direction gives a multivariate sign, and its level sets
produce quantile contours.  Since $R(Y)\sim\nu$ for every $\mathcal L_Y$, the ranks are
distribution-free.  Cyclical monotonicity thus plays the role of monotonicity
on the line.

\citet{CarlierChernozhukovGalichon2016} make this construction conditional.
Vector quantile regression represents a multivariate outcome as
$Y=Q(U,X)$, where the latent rank $U$ has a fixed reference distribution and
$u\mapsto Q(u,x)$ is a gradient of a convex function.  Independence, or the
weaker restriction that $U$ is mean independent of $X$, becomes an additional
linear constraint on the transport problem.  In the scalar case, vector
quantile regression is equivalent to ordinary quantile regression
\citep{KoenkerBassett1978} with an additional monotonicity constraint; under
correct specification, this constraint is automatically satisfied.  The vector
version provides a model of multidimensional heterogeneous effects.

The same construction has direct economic interpretations.
\citet{EkelandGalichonHenry2012} define multivariate comonotonicity and coherent risk
measures through maximal-correlation transport.  \citet{GalichonHenry2012} use
it to extend dual choice theory and inequality measurement to
multiattribute prospects.  Subsequent statistical work studies center-outward
constructions \citep{HallinEtAl2021}, consistency, rates, and nonparametric
testing \citep{GhosalSen2022}, and exactly distribution-free multivariate rank
tests \citep{DebSen2023}.  Optimal transport therefore provides both the ordering needed to
sort multivariate observations and the coupling needed to compare their risks,
inequalities, and conditional distributions.

\subsubsection{Invert}
\budgetnote{300 words}

Discrete-choice estimation runs the behavioral model backward.  A consumer
with taste vector $\varepsilon$ chooses
\[
y(\varepsilon)\in\arg\max_{y\in\mathcal Y}
\{U_y+\varepsilon_y\},
\]
where $U_y$ is product $y$'s systematic utility.  Given the distribution of
tastes and $U$, the model determines market shares; demand inversion starts
from observed shares $s$ and recovers utilities that rationalize them.  The
social-surplus function
\[
G(U)=\mathbb E\left[\max_y\{U_y+\varepsilon_y\}\right]
\]
is convex.  By the Williams--Daly--Zachary theorem
\citep{Williams1977,DalyZachary1978}, the gradient, or the subgradient when $G$
is not differentiable, gives the vector or set of vectors of choice
probabilities.  Demand inversion therefore solves $s\in\partial G(U)$.
\citet{GalichonSalanie2022} introduce the generalized entropy of choice through
the convex conjugate
$G^*(s)=\sup_U\{\sum_y s_yU_y-G(U)\}$, which transforms the surplus function
$G$, expressed in terms of utilities, into a function of choice probabilities.
Demand inversion can then
equivalently be written $U\in\partial G^*(s)$.  When $s$ is rationalized by the choice rule
above,
\[
G^*(s)=-\mathbb E\left[\varepsilon_{y(\varepsilon)}\right].
\]
Thus the generalized entropy is the negative of the expected idiosyncratic
utility generated by the rationalizing choice rule.
\citet{Berry1994,BerryLevinsohnPakes1995} make demand inversion the central step
of differentiated-product demand estimation.

In optimal transport terms, demand inversion is a semi-discrete problem.  The continuous distribution of taste
shocks is transported onto the discrete distribution of product shares.
Systematic utilities are the potentials on the discrete side, while each
product's choice region is the transport cell containing the shocks assigned
to it.  Adjusting $U$ until every cell has its observed probability is
Kantorovich dual optimization.  Logit yields a closed-form entropy, but the
transport formulation applies to much more general heterogeneity.
\citet{ChiongGalichonShum2016} use this convex duality in dynamic discrete-choice
models, where the first-stage inversion becomes a linear program identical to
the Shapley--Shubik assignment problem.

\citet{BonnetGalichonHsiehOHaraShum2022} push the equivalence further
by reading demand inversion as a two-sided matching problem between consumers
and products.  Product capacities are observed market shares and equilibrium
prices are mean utilities.  Stable-matching methods continue to apply when
utility is not additively separable, the demand map is not smooth, or the
inverse is set-valued.

This interpretation requires caution.  Marriage and trade are genuine
two-sided markets; differentiated-product demand is not.  Consumers choose
products, but products do not choose consumers.  The second margin and its
prices are constructed by the analyst to perform the inversion.  Optimal transport is
therefore a tool rather than the economic model itself.  These connections,
together with their extensions to matching, gravity, and data science, are
developed systematically in \citet{Galichon2026}.

\subsubsection{Certify and characterize}
\budgetnote{260 words}

When optimal transport is used as an analytical tool, its main contribution may be neither an
allocation nor an estimator.  It can instead reveal the structure of a
difficult economic problem or provide a certificate that a proposed solution
is globally optimal.  Mechanism design illustrates both roles: incentive
compatibility creates global restrictions, while optimal transport duality and
Beckmann-type formulations translate them into potentials and flows.  Contact
sets identify which types are served and where bunching occurs, while a tight
primal--dual pair certifies that no feasible alternative performs better.

For characterization, optimal transport identifies both the feasible set and its
geometry.  A coupling-existence argument is used by
\citet{EcheniqueNunez2026} to characterize when two ordinal
preferences admit a common canonical cardinalization.  In multidimensional
screening, building on \citet{RochetChone1998},
\citet{MaTrudingerWang2005,FigalliKimMcCann2011} use
optimal transport cross-curvature to characterize when the principal's problem is
convex.  \citet{McCannZhang2024} derive a dual formulation of the
multidimensional adverse-selection problem and use it to reduce the
Rochet--Chon\'e square model to a free-boundary problem.
Thus optimal transport can characterize feasibility and tractability even when
the original problem is not presented as an assignment problem.

For certification, dual variables become proof objects.  Under increasing
differences, \citet{PerezRichetSkreta2026} formulate the design of
falsification-proof allocation rules as the dual of an optimal transport
problem and use this representation to characterize the optimal rule.
\citet{DaskalakisDeckelbaumTzamos2017} express the multi-good monopolist's dual as
``sweep, then transport'': incentive compatibility first reallocates a signed
measure in convex order, after which mass is transported at $\ell_1$ cost.  A
tight primal--dual pair certifies an optimal menu and identifies regions of
exclusion, bunching, or full allocation.  With several bidders, the nonlocal
feasibility constraints require Beckmann flows rather than couplings;
\citet{KolesnikovSandomirskiyTsyvinskiZimin2022} interpret the optimal flow
as a multidimensional generalization of ironed virtual values.

\section{Extensions and open directions}
\budgetnote{700 words}

\subsection{Extensions}
\budgetnote{450 words}

\subsubsection{Equilibrium transport and imperfectly transferable utility}
\label{sec:equilibrium-transport}
\budgetnote{90 words}

Optimal transport takes transferable utility and aggregate surplus maximization
as its benchmark.  Equilibrium transport instead uses pairwise feasibility
conditions to determine which utility combinations a match can sustain.  It
thereby accommodates taxes, bargaining frictions, risk aversion, and
nontransferable utility.  Equilibrium transport builds on decentralized salary
adjustment and gross substitutes \citep{CrawfordKnoer1981,KelsoCrawford1982},
stability \citep{DemangeGale1985,LegrosNewman2007}, and the unified ITU
framework \citep{GalichonKominersWeber2019,GalichonWeber2024}.  Its
equilibrium-flow formulation retains transport structure: potentials become
utilities, and feasibility replaces the dual inequality.  Under
substitutability, this formulation yields existence, comparative statics, and algorithms
\citep{GalichonSamuelsonVernet2024,GalichonSamuelsonVernet2025}.

\subsubsection{Congestion and allocation-dependent interactions}
\budgetnote{70 words}

Optimal transport assigns each route or pair a cost fixed independently of the
allocation.  Congestion breaks this separability: an edge's cost depends on its
total flow, so individual route choices must be determined jointly.
\citet{Wardrop1952} gives the equilibrium condition; when the resulting cost
field is integrable, meaning that it is the gradient of a convex potential, the
equilibrium minimizes that potential \citep{BeckmannMcGuireWinsten1956},
recovering optimization without reducing the problem to ordinary optimal transport.  Peer or
scale effects create analogous feedback in matching, since a pair's payoff
depends on how others match \citep{MourifieSiow2021}.

\subsubsection{Beyond one-to-one and balanced matching}
\budgetnote{130 words}

The bilateral benchmark assigns each unit separately.  With indivisible bundles
and nonadditive valuations, package assignments still admit linear programs,
while integrality determines whether competitive package prices support the
efficient allocation \citep{BikhchandaniOstroy2002}.  This structure also permits
large-scale estimation of random utility over bundles \citep{DiPasqualeWP}.
Team formation with finitely many workers per firm leads to multi-marginal
optimal transport \citep{CarlierEkeland2010,BoermaTsyvinskiZimin2025}.  In
many-to-one production, ordinary capacitated optimal transport suffices when a
firm's surplus is additive across workers.  When output depends on workforce composition,
weak optimal transport makes the firm's payoff depend nonlinearly on its entire
hiring distribution; unnormalized kernels also make firm size endogenous
\citep{ChoneGozlanKramarz2023}.  This extension can represent and compute the
planner's allocation with complementarities.  Such models can also be brought
to data in large-market settings to jointly estimate worker--firm sorting and
coworker complementarities \citep{Corblet2026}.  In a large-firm context,
within-firm additive aggregation of workers guarantees the existence of a
competitive equilibrium that decentralizes the optimal matching
\citep{ChoneKramarzNordstromSkans2025}.  By contrast, in models with discrete
types and indivisibilities, weak optimal transport does not by itself guarantee
the existence of a stable decentralized many-to-one matching; substitutability
or a separate equilibrium argument is needed.

\citet{KolotilinWolitzky2026} instead define an assignment as a measure on
$\mathcal P(\mathcal X)\times\mathcal Y$ whose $\mathcal Y$-marginal is the
distribution of firms and whose average $\mathcal X$-measure is the worker
distribution.  Because each firm hires a unit mass, this is
the normalized counterpart of the endogenous-size formulation above.  The
convexified assignment space yields competitive-equilibrium and core existence
without concavity or substitutability.  Additive surplus recovers standard
optimal transport, while homogeneous firms recover Bayesian persuasion; the
continuum-workforce constraint becomes Bayes plausibility rather than finitely
many marginals.  With mean-measurable production, increasing differences and
decreasing returns yield a concave maximization under a majorization constraint,
isolating a sorting--equality tradeoff and delivering comparative statics.  Thus
the formulation yields equilibrium insight not by restoring substitutability,
but by enlarging and convexifying the assignment space.

\subsubsection{From static to dynamic transport}
\budgetnote{80 words}

Not every dynamic formulation of optimal transport is a dynamic economic model.  The dynamic
formulation of Benamou and Brenier replaces couplings with time-indexed flows
\citep{BenamouBrenier2000}; Schr\"odinger bridges describe stochastic evolution
between fixed endpoints.  Economic dynamics arise when current
allocations change future types and opportunities.  In repeated matching,
today's matches affect tomorrow's populations and continuation utilities, making
equilibrium a fixed point across transport problems
\citep{CorbletFoxGalichon2025}.  Online transport instead treats sequential
arrivals and irreversible assignments, introducing the option value of waiting.

\subsubsection{Optimal transport under additional economic constraints}
\budgetnote{80 words}

Classical optimal transport restricts a coupling only through its marginals;
economic problems often replace or supplement these with other linear
constraints.  Information design asks a sender to choose what agents learn
about an unknown state.  A feasible signal induces a distribution of posterior
beliefs whose mean equals the prior, a requirement known as Bayes plausibility;
obedience additionally requires recommended actions to be optimal.  These
conditions replace the second marginal restriction of classical optimal
transport.
\citet{ArieliBabichenkoSandomirskiy2023} show that
state-conditional belief distributions in multi-receiver persuasion are
feasible precisely when each receiver's marginals are separately feasible;
choosing their correlation then becomes a multi-marginal optimal transport
problem.  \citet{MalamudSchrimpf2022} use martingale optimal transport geometry
to obtain deterministic information policies on low-dimensional manifolds.
Working in the framework introduced by
\citet{KamenicaGentzkow2011}, \citet{Kolotilin2018} and
\citet{DworczakMartini2019} obtained early strong-duality results without
explicitly invoking optimal transport.  \citet{DworczakKolotilin2024} extend
strong duality to information design, while \citet{KolotilinCorraoWolitzky2025}
make the optimal transport connection explicit and obtain a Brenier-type
characterization.  Under their one-dimensional-action and single-peakedness
assumptions, an optimal signal has posteriors supported on at most two states.

Martingale restrictions in finance provide another canonical example: among
couplings satisfying $\mathbb E[Y\mid X]=X$, martingale optimal transport yields
model-independent no-arbitrage bounds
\citep{BeiglbockHenryLaborderePenkner2013,
GalichonHenryLabordereTouzi2014}.  The same restriction arises under rational
expectations: subjective beliefs $\psi$ and realizations $Y$ must admit a
coupling satisfying $\mathbb E[Y\mid\psi]=\psi$.
\citet{DHaultfoeuilleGaillacMaurel2021} use Strassen's theorem to characterize
this requirement in terms of convex order.  Optimal transport also yields sharp portfolio-risk
bounds and informs risk measures and optimal risk allocation
\citep{Ruschendorf2013}.  Moment, capacity, and incentive restrictions in
allocation and mechanism design are analogous.  Such constraints may
nevertheless destroy integrality or Monge structure, producing fractional or
multi-marginal plans.

\subsection{Open directions}
\budgetnote{250 words}

\subsubsection{How should optimal transport allocate while learning?}
\budgetnote{125 words}

Static optimal transport treats the marginals, costs, and available information as given.  In
online markets, today's allocation changes tomorrow's types, opportunities, and
observations.  Ratings, queues, rankings, and adjustment paths are both
coordination devices and endogenous data.  Dynamic and online transport must
therefore determine how a mechanism should trade current welfare against
informative experimentation, while accounting for strategic participation and
the evolution of the market itself.  This calls for economic versions of online
optimal transport in which the surplus and even the marginals are partly unknown, are learned
from endogenous observations, and evolve in response to past assignments.  The
resulting welfare-information frontier may become as central to adaptive
allocation as the efficiency frontier is to static transport.

\subsubsection{Can multiple markets identify unobserved heterogeneity?}
\budgetnote{125 words}

In empirical matching and choice models, unobserved heterogeneity generates the
regularizer that smooths the transport problem.  Its distribution is usually
specified in advance, most often through a logit assumption.  A single market
generally cannot identify both this distribution and the systematic surplus:
changes in one can be offset by changes in the other.
\citet{FoxYangHsu2018} study the related identification of unobserved
complementarities from matching patterns and show how information on unmatched
agents can restore identification that observed matches alone leave incomplete.
Multiple markets offer a possible way forward.  If the distribution of heterogeneity is stable while
population margins, prices, or observable environments vary, the response of
the equilibrium coupling may reveal the regularizer itself.  The open questions
are what cross-market variation suffices, how to distinguish taste dispersion
from surplus curvature, and how to estimate the heterogeneity distribution
flexibly while preserving the convexity and computational tractability that make
regularized optimal transport useful.

\section*{Acknowledgments}

We thank Manuel Arellano, Job Boerma, Philippe Chon{\'e}, Xavier D'Haultf{\oe}uille, Piotr
Dworczak, Federico Echenique, Christophe Gaillac, Florian Gunsilius, Anton Kolotilin, Lihua Lei,
Ilse Lindenlaub, Matthew Shum, Vincent Starck, Keita Sunada, Aleh Tsyvinski, and Zhenyuan Zhang
for helpful comments and suggestions.  We are especially grateful to Bernard Salani\'e for his
extensive and detailed comments, and to Larry Samuelson, the editor, for his
guidance and encouragement.  We also thank Stefania Uccheddu for editorial
assistance.

\bibliographystyle{ar-style1}
\bibliography{references}

\end{document}